\documentclass[prl,twocolumn,showpacs,preprintnumbers,amsmath,floatfix]{revtex4}
\usepackage{amssymb}

\usepackage{graphicx}

\begin{document}

\title{Two pressure-induced transitions in TiOCl: Mott insulator to anisotropic metal}
\author{Yu-Zhong Zhang, Harald O. Jeschke, and Roser Valent\'\i}
\affiliation{Institut f\"ur Theoretische Physik,
Goethe-Universit\"at Frankfurt, Max-von-Laue-Stra{\ss}e 1, 60438
Frankfurt am Main, Germany}
\date{\today}

\begin{abstract}
Using Car-Parrinello molecular dynamics calculations we investigate
the behavior of the low-dimensional multiorbital Mott insulator TiOCl
under pressure. We show that the system undergoes {\it two}
consecutive phase transitions, first at $P_\text{c}$ from a
Mott-insulator to a metallic phase in the $ab$ plane with a strong
Ti-Ti dimerization along $b$.  At a pressure $P'_\text{c} >
P_\text{c}$ the dimerization disappears and the system behaves as a
uniform metal. This second transition has not yet been reported
experimentally. We show that the insulator-to-metal transition at
$P_\text{c}$ is driven by the widening of the bandwidth rather than
structural changes or reduction of crystal field splittings and it
shows a redistribution of the electronic occupation within the
$t_{2g}$ bands.  Our computed pressure-dependent lattice parameters
are consistent with experimental observations and the existing
controversy on the change of crystal symmetry at high pressures is
discussed.

\end{abstract}


\maketitle

Understanding the Mott insulator-to-metal transition in strongly
correlated systems is a central issue in condensed matter physics due
to the striking phenomena, like unconventional superconductivity,
emerging at the vicinity of the
transition~\cite{Imada,SCorganic}. Application of
chemical~\cite{NiS,RNiO3} or hydrostatic pressure~\cite{V2O3} to a
Mott insulator is an ideal way to trigger the Mott transition, as the
one-electron bandwidth ($W$) is directly controlled by pressure,
leading to the change of electron correlation strength ($U/W$) with
$U$ the effective on-site Coulomb interaction. However,
pressure-induced structural changes, as present in FeO~\cite{FeO} or
MnO~\cite{MnO}, and orbital repopulation in multiorbital systems
increase the degree of complexity of the transition and make it
difficult to identify as a Mott transition.

Recently, TiOCl, a layered Mott insulator at room temperature, has
attracted considerable
interest~\cite{Seidel,Shaz,Rueckamp,Abel,Hoinkis} due to its
unconventional transitions to a spin-Peierls state through a
zero-field structural incommensurate phase. This phase has been
ascribed to competing large magnetoelastic couplings with
ferromagnetic interchain frustration~\cite{ZhangJeschkeValenti}. The
idea of inducing a metallic state or even superconductivity by
application of pressure has recently been tried out by some
experimental
groups~\cite{Kuntscher,Forthaus,Blanco-Canosa,Kuntscher1}. Kuntscher
{\it et al.}~\cite{Kuntscher,Kuntscher1} report from optical
measurements and X-ray diffraction analysis the observation of an
insulator-to-metal transition at 16~GPa accompanied by a structural
symmetry change.  Forthaus {\it et al.}~\cite{Forthaus}, on the
contrary, neither observe any indication of a metallic state up to
pressures of 24~GPa from transport measurements nor do they identify
from X-ray diffraction analysis changes on the crystal symmetry at
16~GPa. Recently, Blanco-Canosa {\it et al.}~\cite{Blanco-Canosa}
found a dimerized insulating state at high pressure by X-ray
diffraction and magnetization measurements combined with {\it ab
initio} calculations.

In view of the existing controversy, we present {\it ab initio}
Car-Parrinello molecular dynamics results for TiOCl under pressure,
where both lattice dynamics and electronic properties are treated on
the same footing. Correlation effects are considered within the
generalized gradient approximation GGA+U functional.  We find, in
agreement with Refs.~\onlinecite{Kuntscher} and
\onlinecite{Kuntscher1}, that TiOCl undergoes an insulator-to-metal
transition at a critical pressure $P=P_\text{c}$ and the transition
is accompanied by a lattice dimerization along $b$ with a change
from orthorhombic $Pmmn$ to monoclinic $P2_1/m$ symmetry.  This
structural change is consistent with
Ref.~\onlinecite{Blanco-Canosa}.  Moreover, our results show that
the system undergoes a second phase transition from a dimerized
($P2_1/m$ symmetry) to a uniform (undimerized) metal ($Pmmn$
symmetry) at a pressure $P=P_\text{c}'> P_\text{c}$.  This
transition has not yet been reported experimentally.

The microscopic analysis of the insulator-to-metal transition at $P
= P_\text{c}$ reveals that this transition happens in the $ab$
bilayers with closure of the gap at the Fermi level, while the
conductivity along $c$ may be hindered by the large van der Waals
gap between different bilayers, reminiscent of the anisotropic
conductivity in cuprates upon doping~\cite{Fujii}. The transition is
driven by the widening of the Ti $3d$ bands rather than a reduction
of the crystal field splittings and we find orbital repopulation of
the electronic density in the $t_{2g}$ orbitals. The concomitant
dimerization along $b$ is attributed to the pressure-enhanced
hybridization.  We show that both phase transitions at $P_\text{c}$
and $P'_\text{c}$ are of first order.  The calculated
pressure-induced changes of lattice parameters are in good agreement
with the experimental results~\cite{Forthaus,Kuntscher1} and we can
explain their strong anisotropic properties as well as their abrupt
jump behavior (contraction-expansion) at the transitions.

For our calculations, we employed the
Car-Parrinello~\cite{CarParrinello} projector-augmented wave (CP-PAW)
method~\cite{Bloechl} and studied the pressure effects on the
electronic and lattice variables with the Parrinello-Rahman fictive
Lagrangian approach~\cite{Rahman}. The great advantage of this
procedure is the ability to deal with crystals with anisotropic
compressibilities along different directions as is the case for
TiOCl. In order to have the correct insulating behavior at ambient
pressure, we employed the GGA+U functional. It has already been
shown~\cite{Pisani,ZhangJeschkeValenti} that strong correlations are
also important for the correct description of the lattice structure.
In fact, the lattice parameters and the distances and angles between
atoms obtained from CP-PAW with GGA+U at ambient pressure are more
consistent with the experimental results than GGA or local density
approximation (LDA) determined structures. We performed carefully
converged calculations with high energy cutoffs of 45 Ry and 180 Ry
for the wave functions and charge density expansion, respectively.  A
value of $U=1.65$~eV with which correct spin exchange along $b$ is
reproduced~\cite{ZhangJeschkeValenti} was used for the GGA+U
calculations. The relation between the $U$ value used in different
codes is discussed in ref.~\cite{ZhangJeschkeValenti}.

\begin{figure}[tb]
\includegraphics[angle=-90,width=0.45\textwidth]{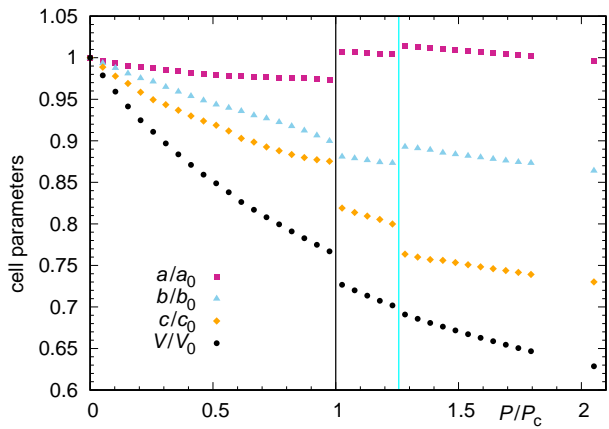}
\caption{(Color online) Pressure dependence of lattice parameters
and volume. Pressure is normalized to the critical pressure for the
insulator-to-metal transition and lattice parameters are normalized
to their values at ambient pressure. The compressibility along $a$,
$b$ and $c$ is strongly anisotropic. The existence of two phase
transitions is recognizable by jumps in all quantities.}
\label{fig:Lattice}
\end{figure}

In Fig.~\ref{fig:Lattice} we present the calculated change of
lattice parameters as a function of applied external pressure.  The
lattice parameters at each pressure are obtained by fully relaxing
the lattice structure without any symmetry constraints. Although a
spin ordered state has to be imposed within the GGA+U functional,
the results are completely consistent with the
experiments~\cite{Forthaus,Kuntscher1}. We observe that the largest
compression happens along $c$ since the interaction between
neighboring bilayers is of van der Waals type. Along $b$, which is
the direction of the Ti chains with strong antiferromagnetic (AF)
interaction $J_b$~\cite{Seidel}, a noticeable shrinkage is observed
as pressure increases. The magnitude of the shrinkage depends on the
choice of the spin configuration set in the GGA+U calculation.  The
experimental results are reproduced when we set AF order along $b$
while this direction becomes incompressible if the spin
configuration is set to be ferromagnetic (FM). This is due to the
fact that the Pauli principle prevents two electrons with the same
spin to occupy the same bonding state with low energy. Therefore, if
the volume along $b$ is reduced in a FM spin configuration, the
total energy will increase rapidly due to occupation of an
antibonding state as two Ti ions come closer.  For an AF
configuration along $b$, the Pauli principle doesn't affect the
energy and the system becomes compressible along $b$ as
observed~\cite{Forthaus}.

In contrast, TiOCl remains almost incompressible along $a$ regardless
of the spin configurations we chose along this direction. This is due
to the fact that the Ti-Ti interaction along $a$ is of
Ti$\,d_{x^2-y^2}$-O$\,p$-Ti$\,d_{x^2-y^2}$ superexchange nature, and
not direct exchange as along $b$. In this case spin order is
irrelevant for compressibility. Throughout this work, we adopted AF
order along $b$ and FM along $a$ and $c$~\cite{ZhangJeschkeValenti}.

At $P=P_\text{c}$ (black vertical line in Fig.~\ref{fig:Lattice}) all
lattice parameters show an abrupt jump. While $b$ and $c$ contract,
the $a$ direction expands, as experimental observations seem to
indicate~\cite{Kuntscher2}. We find a second jump of the lattice
parameters at $P=P'_\text{c} > P_\text{c}$ (blue vertical line in
Fig.~\ref{fig:Lattice}) that defines a second phase transition. At
this pressure, both $a$ and $b$ directions show an abrupt expansion.
In order to understand the nature of these phase transitions, we now
analyze the density of states (DOS) and change of symmetry.

\begin{figure}[tb]
\includegraphics[width=0.45\textwidth]{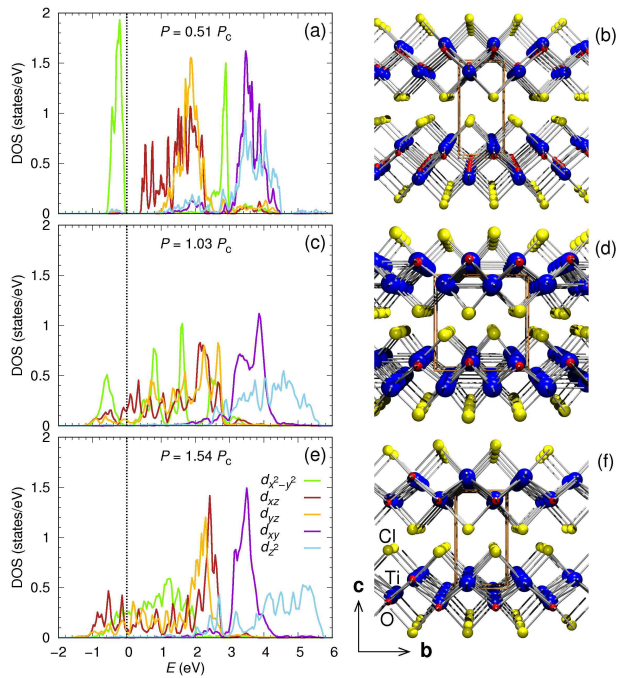}
\caption{(Color online) (a), (c), and (e) are the partial density of
states at three different values of pressure in different phases. In
the insulating state, only $d_{x^2-y^2}$ orbital is occupied while all
the $t_{2g}$ bands are occupied in the metallic regions. In the
dimerized metallic phase, a gap still opens in $d_{x^2-y^2}$
orbital. (b), (d), and (f) is the structure at these three different
values of pressure. The structure in the dimerized metal is quite
different from the other two. In the metallic phase, all the O atoms
behind Ti atoms are invisible due to the flatness of the Ti-O-Ti
angle.} \label{fig:DOS}
\end{figure}

In Fig.~ \ref{fig:DOS} we show the Ti $d$ partial DOS in the range
of energies between -2~eV and 6~eV at pressures in the vicinity of
$P_\text{c}$ and $P'_\text{c}$.  We chose the local coordinate
system as $x\parallel b$, $y\parallel c$ and $z\parallel a$, which
diagonalizes the Hamiltonian with $d_{x^2-y^2}$, $d_{xz}$, $d_{yz}$
forming the $t_{2g}$ bands and $d_{xy}$, $d_{z^2}$ the $e_g$ bands.
In the insulating phase at $P/P_\text{c}=0.51$ (see
Fig.~\ref{fig:DOS}~(a) and (b)), a gap is present in the DOS,
separating the occupied $d_{x^2-y^2}$ state from the unoccupied
$d_{yz}$ and $d_{xz}$ states. The whole character of the DOS and
structure is similar to that at ambient pressure for temperatures
above the spin-Peierls transition~\cite{Seidel,Hoinkis} except for a
reduction of the gap and the width of the $d$ bands. At
$P/P_\text{c}=1.03$, right above the transition (see
Fig.~\ref{fig:DOS}~(c) and (d)), the gap disappears and the $d$
states are suddenly rearranged. The Fermi level crosses the $d_{yz}$
and $d_{xz}$ states while a gap within $d_{x^2-y^2}$ (orbital in the
$bc$ plane) still remains. These results indicate an
insulator-to-metal transition at $P_\text{c}$.  Moreover, the abrupt
expansion along $a$ can be explained by the occupation of the
$d_{yz}$ and $d_{xz}$ orbitals which strongly overlap between
neighboring Ti ions along $a$. Direct repulsive interactions between
same-spin electrons in these orbitals based on the Pauli principle
lead to a sudden expansion. We also observe a sudden contraction of
the structure along $b$ and a dimerized pattern, as shown in
Fig.~\ref{fig:DOS}~(d), which is due to the occurrence of tighter
Ti-Ti dimer states between neighboring sites along $b$ and explains
the gap in the $d_{x^2-y^2}$ partial DOS. The obtained
insulator-to-metal transition is in agreement with optical
experiments~\cite{Kuntscher} but contradicts electrical resistivity
measurements~\cite{Forthaus}.  TiOCl is a strongly anisotropic
crystal with a large van der Waals gap along $c$. If the resistivity
measurements have been done along $c$, the predominance of this gap
is to be expected. We observe from our calculations that the
metalicity is in the $ab$ bilayers~\cite{Kuntscher}.

In Fig.~\ref{fig:DOS}~(e) we show the DOS at $P= 1.54 P_\text{c}$
right above the transition at $P = P'_\text{c}$. The gap in
$d_{x^2-y^2}$ as well as the structural dimerization along
$b$ disappear (Fig.~\ref{fig:DOS}~(d)). The system shows a uniform metallic
phase which has not yet been established experimentally. The
abrupt increase of occupation numbers in $d_{yz}$ and $d_{xz}$
orbitals at $P = P'_\text{c}$ accounts for another sudden expansion in
$a$, and the disappearance of the dimer states along $b$ for the
abrupt expansion of $b$.

\begin{figure}[tb]
\includegraphics[angle=-90,width=0.45\textwidth]{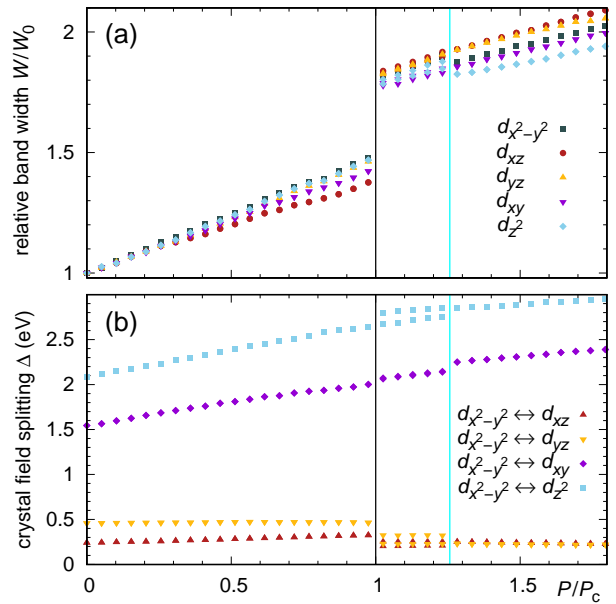}
\caption{(Color online) (a) Bandwidth of all  five $d$ states as a
function of applied pressure. Pressure is normalized to the critical
pressure, bandwidth to the value at ambient pressure. (b) Crystal
field splittings from $d_{x^2-y^2}$ to the other $d$ orbitals as a
function of pressure. Lower symmetry in the dimerized metallic phase
produces two values instead of one for both quantities.}
\label{fig:BWCS}
\end{figure}

In Fig.~\ref{fig:BWCS} we present the pressure-induced changes of
the $d$-state bandwidths as well as the crystal field splittings
between the lowest $d$-band ($d_{x^2-y^2}$) and all other $t_{2g}$
and $e_g$ bands. We observe within the three phases that the
bandwidths of all $d$ states widen monotonically with pressure
(Fig.~\ref{fig:BWCS}~(a)), while the crystal field splittings show
two different trends (Fig.~\ref{fig:BWCS} (b)). The orbital
excitations from $d_{x^2-y^2}$ to the $e_g$ orbitals
 increase monotonically with pressure, in agreement with
experiment (for the optically allowed excitation $d_{x^2-y^2}$ to
$d_{xy}$)~\cite{Kuntscher}. Meanwhile, the excitations to the
$t_{2g}$ orbitals depend very little on pressure. From these results
we conclude that the insulator-to-metal transition is mainly
controlled by an increase of bandwidths rather than
by crystal field
splitting.  However, the phase transition is characterized by a
redistribution of electronic occupation among all three
$t_{2g}$ bands. Thus, it is not a typical one-band Mott transition
where electron transfer occurs only within one band.  Moreover, in
agreement with the sudden changes of lattice parameters at pressures
$P=P_\text{c}$ and $P=P'_\text{c}$, bandwidths and crystal field
splittings also show abrupt jumps at the phase transitions. The
cooperative effect between variations of electronic bandwidths and
changes of crystal field splittings by lattice distortions may be
the mechanism responsible for the first order phase transitions.

\begin{figure}[tb]
\includegraphics[angle=-90,width=0.45\textwidth]{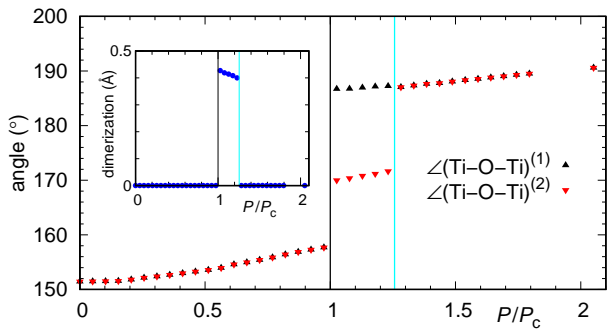}
\caption{(Color online)
 angle between Ti-O-Ti as a function of pressure
normalized by the critical pressure.
 The inset  shows
the difference between long and short Ti-Ti bonds along $b$.}
\label{fig:LDD}
\end{figure}

The analysis of Ti-O, Ti-Cl distances as well as
 Ti-O-Ti angles as a function of pressure shows  that, as  expected, the
distances between atoms in the insulating phase
decrease monotonically with increasing pressure. Nonetheless, the distances
between Ti-O along $a$  increase suddenly
  at pressures $P_\text{c}$ and $P'_\text{c}$  due to the expansion along
this direction at the phase transitions. The
strongly distorted octahedron [TiO$_4$Cl$_2$] at ambient pressure is
modified towards a regular octahedron by a flattening of
the Ti-O-Ti angle with increasing pressure (Fig.~\ref{fig:LDD}).
At $P > P'_\text{c}$, the angles become slightly larger than 180 degree.

A subject of controversy has been the crystal symmetry for pressures above
$P_\text{c}$~\cite{Kuntscher,Forthaus,Kuntscher1}. We have
identified the calculated structure as orthorhombic space group
$Pmmn$ (59) for pressures $P < P_\text{c}$ and $P > P'_\text{c}$ and
monoclinic space group $P2_1/m$ (11) for pressures $P_\text{c} < P <
P'_\text{c}$.  In order to confirm that the restoration of the high
symmetry $Pmmn$ at pressures $P > P'_\text{c}$ is independent of the
choice of unit cell and spin order, we also performed calculations
with different spin configurations (AF/FM) for different supercells
and artificially disordered structures at pressures $P>P'_c$. The
high symmetry  $Pmmn$ is recovered in all cases. These
investigations indicate that the lattice first undergoes a
transition from orthorhombic to monoclinic structure as dimerization
sets in, which is consistent with Ref. \onlinecite{Blanco-Canosa},
and under further pressure increase a transition back to the
orthorhombic symmetry occurs as the dimerization vanishes due to
nonlinearities in the elastic response. The abrupt onset and
disappearance of dimerization shown in the inset of
Fig.~\ref{fig:LDD} indicates that both phase transitions are of
first order. Such complicated structural phase transitions and
 high lattice anisotropy
 could be the origin of the difficulties in determining
 the lattice structure from X-ray diffraction data, leading to
different conclusions from
experiments~\cite{Kuntscher,Forthaus,Blanco-Canosa,Kuntscher1}.

Finally, we note that the phase transitions in the CP-PAW
calculations with the GGA+U functional occur at higher pressures
($39\pm 1$~GPa) than the experimental pressures ($\approx 16$~GPa).
Such deviations are attributed to the artificial long-range AF order
along $b$ we imposed, which is used to mimic the strong correlation
in the GGA+U approach. However, as is well-known, only short-range
AF fluctuations exist in a Mott insulator. The situation can be
clearer if we resort to the phase diagram shown in
Ref.~\cite{Kotliar}. It is found that one has to apply much higher
pressure to drive an AF insulator to a Fermi liquid than to drive a
Mott insulator to a metal. To remedy this shift in the transition
pressure, relaxation with a yet to be developed constant pressure
{\it ab initio} molecular dynamics LDA+DMFT method would be
required.

In conclusion, our {\it ab initio} molecular dynamics study of TiOCl
under pressure shows that this system undergoes two phase transitions,
the first one from an insulator to an anisotropic metal accompanied
with a strong structural Ti-Ti dimerization along $b$, and a second
one from the dimerized to a uniform metallic state.  Our results
indicate that the Mott insulator-to-metal transition is mainly driven
by bandwidth widening leading to an abrupt change of the electronic
structure and orbital population and inducing a structural change.  We
propose resistivity experiments contacting the $ab$ planes in
order to settle the question of insulator or metal as well as
analysis at higher pressures to confirm our prediction of a second
phase transition. Due to the orbital repopulation and the structural
change, which leads to a dimensional crossover from quasi-one
dimensional to quasi-two dimensional system at $P > P'_\text{c}$, the
question of possible superconductivity is still open.

We acknowledge useful discussions with P. Bl\"ochl, C. A. Kuntscher,
R. Claessen, M. Sing, M. Abd-Elmeguid, T. Saha-Dasgupta and C. Gros.
We thank the Deutsche Forschungsgemeinschaft for financial support
through the TRR/SFB~49 and Emmy Noether programs and we acknowledge
support by the Frankfurt Center for Scientific Computing.


\begin{thebibliography}{99}
\bibitem{Imada}
M. Imada {\it et al.}, Rev. Mod. Phys. {\bf 70}, 1039 (1998).

\bibitem{SCorganic}
Y. Kurosaki {\it et al.}, Phys. Rev. Lett. {\bf 95}, 177001 (2005).

\bibitem{NiS}
K. Iwaya {\it et al.}, Phys. Rev. B {\bf 70}, 161103(R) (2004).

\bibitem{RNiO3}
J. B. Torrance {\it et al.}, Phys. Rev. B {\bf 45}, 8209 (1992).

\bibitem{V2O3}
D. B. McWhan {\it et al.}, Phys. Rev. B {\bf 7}, 1920 (1973).

\bibitem{FeO}
Z. Fang {\it et al.}, Phys. Rev. Lett. {\bf 81}, 1027 (1998).

\bibitem{MnO}
J. Kunes {\it et al.}, Nature Materials {\bf 7}, 198 (2008).

\bibitem{Seidel}  A. Seidel {\it et al.},
Phys. Rev. B \textbf{67}, 020405(R) (2003).
T. Saha-Dasgupta {\it et al.}, Europhys. Lett. {\bf 67},
63 (2004).

\bibitem{Shaz}
M. Shaz {\it et al.}, Phys. Rev. B {\bf 71}, 100405(R) (2005); S.
van Smaalen {\it et al.}, Phys. Rev. B {\bf 72}, 020105(R) (2005).

\bibitem{Rueckamp}
R. R\"{u}ckamp {\it et al.}, Phys. Rev. Lett. {\bf 95}, 097203
(2005).

\bibitem{Abel}
E. T.  Abel {\it et al.}, Phys. Rev. B {\bf 76}, 214304 (2007).

\bibitem{Hoinkis}
M. Hoinkis, {\it et al.}, Phys. Rev. B {\bf 72}, 125127 (2005).

\bibitem{ZhangJeschkeValenti}
Y. Z. Zhang, {\it et al.}, arXiv:0804.3342 (2008).

\bibitem{Kuntscher}
C. A. Kuntscher {\it et al.}, Phys. Rev. B {\bf 74}, 184402 (2006).

\bibitem{Forthaus}
M. Forthaus {\it et al.}, Phys. Rev. B {\bf 77}, 165121 (2008).

\bibitem{Blanco-Canosa}
S. Blanco-Canosa {\it et al.}, arXiv:0806.0230v1 (2008).

\bibitem{Kuntscher1}
C. A. Kuntscher {\it et al.}, Phys. Rev B {\bf 78}, 035106  (2008).

\bibitem{Fujii} T.
 Fujii {\it et al.}, Phys. Rev. B {\bf 66}, 024507 (2002).

\bibitem{CarParrinello}
R. Car, M. Parrinello, Phys. Rev. Lett. {\bf 55}, 2471 (1985).

\bibitem{Bloechl}
P. E. Bl\"ochl, Phys. Rev. B {\bf 50}, 17953 (1994).

\bibitem{Rahman}
M. Parrinello, A. Rahman, Phys. Rev. Lett. {\bf 45}, 1196 (1980).

\bibitem{Pisani}
L. Pisani, R. Valent\'\i, Phys. Rev. B {\bf 71}, 180409(R) (2005).

\bibitem{Kuntscher2}
C. A. Kuntscher, private communication.

\bibitem{Kotliar}
G. Kotliar, Science {\bf 302}, 67 (2003).

\end{thebibliography}
\end{document}